\documentclass[english,12pt]{article}
\usepackage{fullpage}
\usepackage{authblk}

\usepackage[pdftex]{graphicx} 

\usepackage[pdfpagemode=None,colorlinks=true,urlcolor=red,%
linkcolor=blue,citecolor=black,pdfstartview=FitH]{hyperref}

\usepackage{url}

\usepackage{amsfonts,amssymb,amsbsy}

\usepackage{amsmath}
\usepackage{braket}
\usepackage{listings}
\usepackage{booktabs}
\usepackage{caption}
\usepackage{subcaption}

\title{A multi-scale code for flexible hybrid simulations}

\author[1]{L. Leukkunen}
\author[2]{T. Verho}
\author[1]{O. Lopez-Acevedo}
\affil[1]{COMP Centre of Excellence, Department of Applied Physics, 
Aalto University, P.O. Box 11100, 00076 Aalto, Finland}
\affil[2] {Molecular Materials, Department of Applied Physics, 
Aalto University, P.O. Box 11100, 00076 Aalto, Finland}

\begin{document}
\maketitle

\begin{abstract}

Multi-scale computer simulations combine the computationally efficient classical
algorithms with more expensive but also more accurate ab-initio quantum
mechanical algorithms. This work describes one implementation of
multi-scale computations using the Atomistic Simulation Environment (ASE). This
 implementation can mix classical codes like LAMMPS and the Density Functional 
Theory-based GPAW. Any combination of codes linked via the ASE interface 
however can be mixed. We also introduce a
 framework to easily add classical force fields calculators for ASE using 
LAMMPS, which also allows harnessing the full performance of classical-only 
molecular dynamics. Our work makes it possible to combine different simulation 
codes, quantum mechanical or classical, with great ease and minimal coding 
effort.
\end{abstract}

\addcontentsline{toc}{section}{Contents}


\pagenumbering{arabic}

\section{Introduction}
\subsection{Multiscale Computations}
Computer simulations are a valuable tool in modern molecular physics research.
They can provide insight into dynamics of a complex system which can be hard
to attain through actual experimentation and they are able to deliver results
in scenarios that can be hopelessly beyond capabilities of current analytical
methods.

Despite its significant potential, computational physics faces challenges in
practice arising from the computational cost associated with many high precision
algorithms. This leads to continuous balancing of the desire for accuracy with
the cost of reaching it. There is a large performance gap between ab initio DFT
algorithms and the computationally cheap but quite system specific classical
force field algorithms. This has created a need for a mechanism that would allow
bridging this chasm in order to simulate systems with thousands of atoms while
retaining the flexibility of modeling quantum mechanical behavior that arises,
without having to first spend significant effort in tuning the classical force
fields to work in the particular system under scrutiny.

There can be considered to be two main approaches to a solution to this. First
method attempts to develop and improve fast classical methods to incorporate
behavior arising from quantum mechanical features. Reactive force fields are
one such example of this class of solutions. The second path is the attempt to
combine two or more algorithms into a single uniform multi-scale simulation. An
example of this approach is ONIOM \cite{oniom} developed by Gaussian. Both
approaches have their strengths and weaknesses and currently no universal answer
exists. Multi-scale setups tend to lead to problems at the interface region
between the calculation methods, while reactive force fields face difficulties
coping with systems that fall outside the setups for which they were trained
for.
\subsection{Atomistic Simulation Environment}
Atomistic Simulation Environment (ASE)\cite{ase} is an open source
software framework designed to perform simulations using a number of algorithm
implementations. These algorithms are called calculators in ASE. The calculator
interface is defined so that an adhering class must provide functions that
calculate both forces acting on the atoms and the total potential energy of the
system. ASE itself is written in Python programming language, but the actual
calculator implementation can be something else, as long as a Python class
wrapping the implementation is provided.

The force and energy information provided by the calculators is used by
various other ASE modules to perform a range of molecular dynamics operations.
These include structure relaxation along with constant NVE/NVT molecular
dynamics. Structure relaxation is also considered a molecular dynamics
operation even if there is no time progression involved.

Using ASE consists typically of writing small Python scripts that create the
atom configuration and setup the calculators and any molecular dynamics
operations that are to be performed. The framework is designed so that
the central object is the Atoms class, which stores all the arrays describing
the atoms of the system. The calculator is then assigned to this Atoms object,
essentially resulting in a one to one mapping. At present the ASE framework
does not provide any facilities for combining multiple calculators into a
single simulation.
\subsection{GPAW}
GPAW \cite{gpaw2} is a projector-augmented wave (PAW) \cite{PAW1} implementation 
designed to work with ASE. PAW takes account of the
core electrons of atoms as frozen, and computes with soft pseudo valence wave
functions. GPAW supports all the exchange-correlation functionals provided
by the libxc library \cite{libxc}. While GPAW reaches good accuracy, it is
compute intensive, requiring large cluster computers to simulate systems larger
than just tens of atoms. For GPAW a huge simulation is already something with
500 atoms and for that kind of calculation a cluster with thousands of nodes
is needed in order to produce results in a reasonable amount of time.
\subsection{LAMMPS/ReaxFF}
LAMMPS \cite{lammps} is a classical molecular dynamics simulator. It is a mature
framework that supports parallel distributed memory cluster computers. Similar
to ASE, it allows different algorithms to be plugged in through an extension
mechanism. One such algorithm is the ReaxFF \cite{reaxff} developed by Adri van
Duin. ReaxFF has been developed to bridge the gap between ab initio quantum
mechanical algorithms, such as GPAW, and the existing  empirical force field
(EFF) methods. Unlike traditional EFF methods, ReaxFF is capable of describing
reactive systems. As is the nature of LAMMPS as a simulation environment,
ReaxFF produces potentials describing the interactions of atoms. These are
trained using ab initio techniques to fit a particular system, but unlike
some EFF algorithms, there is only one version of each atom type used
in a single simulation.

The performance difference between LAMMPS/ReaxFF and ASE/GPAW is significant.
In a simulation of a system containing 5000 atoms of which 10 are computed
using GPAW, the ReaxFF computations take so little time it is not worth
to even consider parallelizing that part as the GPAW computation of
its tiny subset of atoms completely dominates CPU usage.
\section{ASE/LAMMPS interface}
\subsection{Design of the interface}
Since the standard ASE distribution merely offers rudimentary support for 
LAMMPS that only allows nonbonded interaction, we took upon ourselves to 
improve the interface. Because LAMMPS supports a large number of functional 
forms for force fields, a great many force fields can be used through LAMMPS. 
Therefore, in our design, each LAMMPS force field is a separate calculator 
class in ASE. However, all of them inherit a base class for LAMMPS calculators,
 \texttt{LAMMPSBase}, that handles all communication with LAMMPS. The 
responsibility of the subclasses is to (a) provide the force field parameters 
as a \texttt{FFData} object, (b) handle atom typing, i.e. determine a force 
field specific type for each atom, depending on its chemical environment, and 
(c) possibly assign partial charges to atoms for calculating electrostatic 
interactions. The methods in \texttt{LAMMPSBase} then detect chemical bonds, bond 
angles, dihedrals and improper dihedrals in the system and generate LAMMPS 
input based on the force field parameters provided by the subclass.

The main steps for adding support for a new LAMMPS force field consist 
typically of writing a parser that reads a force parameter file acquired from 
an external source, and enabling automatic atom typing for the force field. To 
make automatic typing as easy as possible, we defined a template syntax for 
defining chemical environments that characterize a given type. For each type, 
a template expression consisting of a single line is given, and to avoid 
ambiguity with multiple matching types, precedence expressions for defining 
priority among types can be used.

In order to offer maximal performance in molecular dynamics simulations that 
do not involve mixing with quantum mechanical calculators, we wrote ASE 
dynamics classes that are specific to the LAMMPS calculators. Their purpose is 
to allow running dynamics inside LAMMPS without needing to execute Python code 
between every timestep. The syntax for using these LAMMPS dynamics classes is 
identical to that when using the standard ASE dynamics classes, but instead of 
executing the LAMMPS calculator timestep by timestep, they instruct the 
calculator to execute LAMMPS molecular dynamics runs with given parameters. 
The run is divided into shorter runs in order to save ASE trajectory snapshots 
along the way, also allowing the use of ``observer'' functions in the same way 
as the standard ASE dynamics classes. In the future, the concept of LAMMPS 
specific dynamics classes can be used to expose the rich feature set of LAMMPS 
to the ASE user in a clean way. This could include robust equilibration methods
 such as replica exchange dynamics.

Even though the present paper is geared towards mixing the LAMMPS calculator with quantum codes, we regard the interface potentially useful for purely classical molecular dynamics simulations as well. In addition to allowing running LAMMPS from ASE, it automatizes many task such as atom typing and generating LAMMPS input parameters, something that the LAMMPS native interface leaves to the user.
\subsection{Using the interface}

We illustrate the usage of the LAMMPS interface with a simple molecular dynamics involving a phenol dimer. The initial geometry is obtained from the s22 data set included in ASE, and we demonstrate two different LAMMPS calculators, ReaxFF and CHARMM general force field for drug-like molecules\cite{charmm}. The code in listing \ref{code:lammps dynamics} runs a 10 picoseconds simulation with first ReaxFF and then CHARMM, and saves the trajectories in two separate files.

\lstset{basicstyle=\footnotesize, frame=single, captionpos=b, language=Python}
\lstset{caption=Molecular dynamics with LAMMPS, label=code:lammps dynamics}
\begin{lstlisting}
from ase.data import s22
from ase import units
from multiasecalc.lammps import ReaxFF, CHARMM
from multiasecalc.lammps.dynamics import LAMMPS_NVT
from multiasecalc.utils import get_datafile

atoms = s22.create_s22_system('Phenol_dimer')
atoms.calc = ReaxFF(get_datafile('ffield.reax'))
dyn = LAMMPS_NVT(atoms, 1*units.fs, 300, trajectory='reax.traj')
dyn.run(10000)
atoms.calc = CHARMM(get_datafile('par_all36_cgenff.prm'))
dyn = LAMMPS_NVT(atoms, 1*units.fs, 300, trajectory='charmm.traj')
dyn.run(10000)
\end{lstlisting}

As an aside, we don't implement any partial charge determination scheme in the CHARMM force field, so the user needs to apply a 
suitable charge equilibration method prior to simulation. In listing \ref{code:lammps dynamics}, we use the build-in charge 
equilibration algorithm in ReaxFF to provide partial charges for the CHARMM simulation.

\section{Force and Energy Mixer}
\subsection{Design of The Mixer}
The easiest way to combine results from two calculators is to simply assign
different atoms to each and then present the combined results. This is
guaranteed to produce errors as the interaction between the two sets of atoms
is not taken into account at all.

To avoid this error, the Mixer employs a strategy based on ONIOM, a method
developed by Gaussian. The basic idea is to calculate the full system using
the cheap less accurate algorithm, then correct this result by adding the
result from calculating the quantum mechanically interesting sub-system with
the QM method. At this point the QM sub-system atoms are calculated twice, so
a further correction is done by removing the same QM sub-system calculated
using the classical method.

This still leaves something to be desired, as the outer edges of the QM
sub-system will have substantial errors introduced in the calculations due to
the arbitrary system boundary. To avoid this, the Mixer uses weights
that can be assigned to atoms for each calculation, controlling how big
contribution it will make to the final result. The weights can either be fixed
to each atom permanently, or calculated dynamically depending for example on
the position of the atom.

It is important to keep in mind that the Mixer itself imposes very few
restrictions on how the forces and energies of the sub-calculators are combined.
It can support any number of calculators, used in any number of force and energy
calculations, with atoms and their weights and coefficients set separately for
each and updated dynamically as the simulation progresses. In this work we
primarily concentrate on a setup where two calculators are used but this in no
way implies that the Mixer could not be used in a more creative way.

\begin{figure}[ht!]
 \centering
 \includegraphics[width=0.3\textwidth]{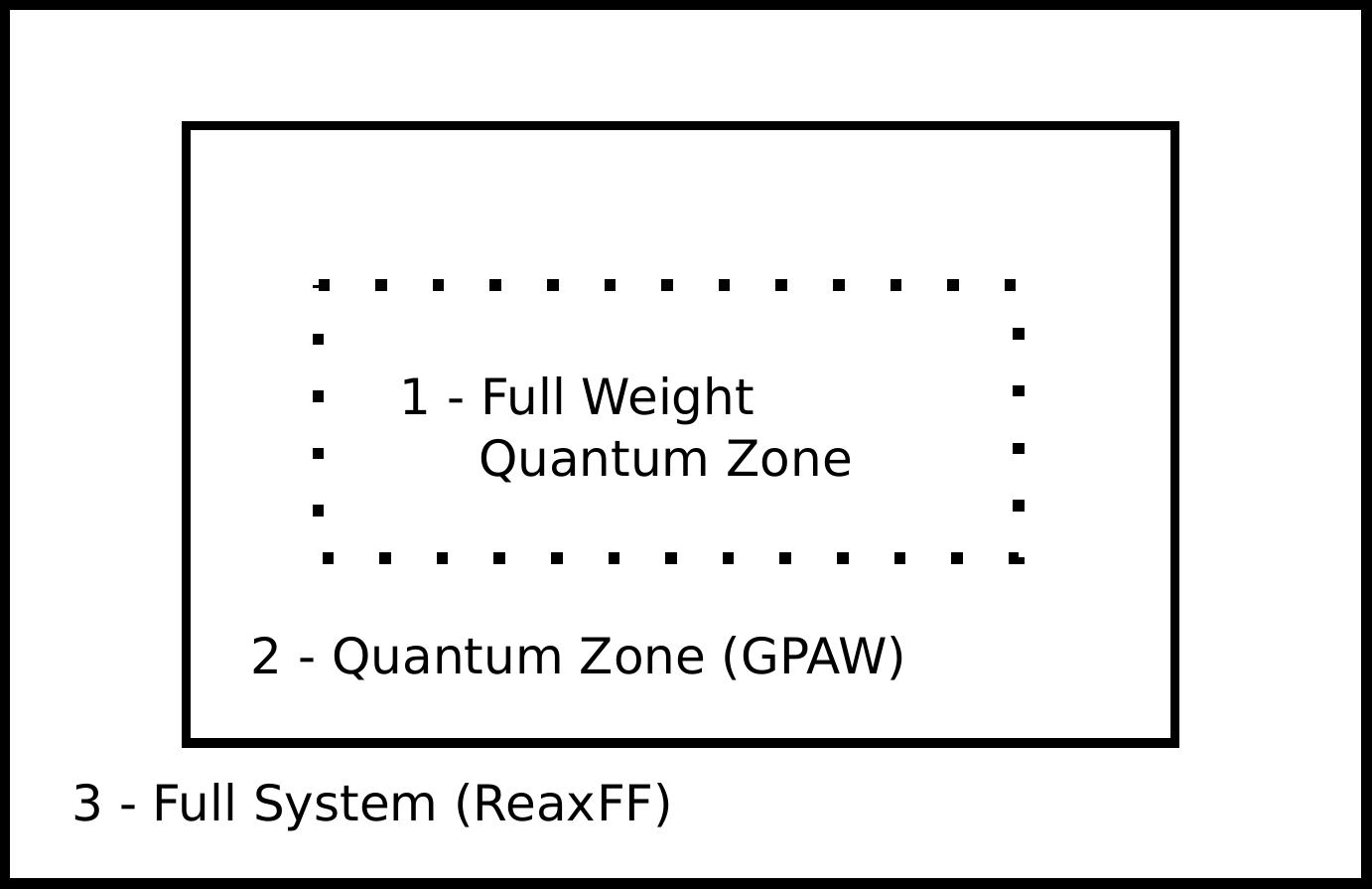}
 \caption{Simple 2D Calculation Region Arrangement}
 \label{fig:calc_box_1}
\end{figure}

For a concrete example, we can consider the simple two calculator configuration
shown in figure \ref{fig:calc_box_1} and have the outer box 3 simulated using a
classical method and a smaller inner box 1 using a QM method. For energies we
are somewhat restricted by how the potential energy is calculated. The ASE
calculators only provide a single potential energy value for the full system
they computed, there is no way to split that up into finer components in a
general way. We will therefore stick to the basic ONIOM Hamiltonian: 
\begin{equation}\label{eq:energies}
H = H_{3C} + \{H_{2Q} - H_{2C}\}.
\end{equation}

Forces are given on a per atom basis by ASE calculators, so we have more
flexibility in calculating the mixed combination. The equation for forces
acting on atom $i$ is:
\begin{equation}\label{eq:forces}
F^i = F_{3C}^i + w^i(F_{2Q}^i - F_{2C}^i).
\end{equation}
The numbers 1, 2, 3 refer to the regions of space so that 1 is the fully quantum
inner box, 2 the inner box + transition region and 3 the full system. Letters
Q and C refer to the calculators used: Q for quantum and C for classical.

This leaves us with a slight discrepancy between the force and energy
calculations but the additional error dampening given by the weights for
forces could help achieve for example better relaxation results. The Mixer
design allows the user to control the weight configuration entirely, affording
complete tailoring of these aspects of the calculation to the problem at hand.
\subsection{Implementation of The Mixer}
The Mixer is designed to present two or more ASE calculators as one towards
the rest of the ASE framework. This is achieved by making the mixer appear as
an ordinary ASE calculator itself. It provides a single set of forces
and a single total potential energy for the given Atoms object. It can
therefore take part in all the molecular dynamics operations ASE supports.
\begin{figure}[ht!]
 \centering
\includegraphics[width=0.7\textwidth]{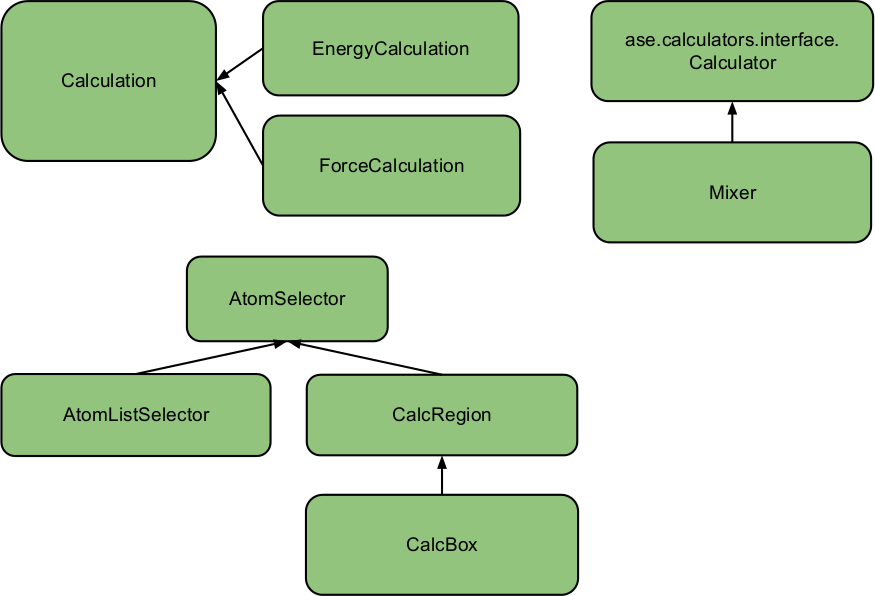}
 \caption{Mixer Class Diagram}
 \label{fig:mixer_class_diag}
\end{figure}

As the Mixer inherits from the ASE interface class for calculators, it meets
the requirements for a normal ASE calculator. It then further extends this
functionality by providing a set of constructor parameters and member methods
for registering a set of force and energy calculations. The main role of the
Mixer class is to simply sum up the contributions from the sub-calculations
using the coefficients given by them. The coefficients are primarily used
to control the sign of the contribution. Almost all actual complexity of the
Mixer concept is separated out to the Calculation and AtomSelector classes.
This has been done to allow the user to completely control the way the
multi-scale calculations are to be combined without having to modify the Mixer
implementation itself.

The Mixer Calculation classes have an ASE calculator and an AtomSelector
assigned to them. The AtomSelector is used to select the correct atoms from
the original full system Atoms object. Two implementations of this selector
functionality are provided. First is AtomListSelector, which filters based on
an atom ID list, and second is the CalcBox, which selects atoms that fall within
its extents. These demonstrate the flexibility provided by the Mixer design
and can be used as a model for how the user can implement custom atom
selectors to meet requirements of a particular simulation. The AtomSelectors
are also responsible for providing the weight used in ForceCalculations.
This means there are two separate factors affecting the computation of forces.
The first is the weight that AtomSelector assigns, and second is the
coefficient provided by the ForceCalculation. This may seem slightly strange
at first, but this design allows re-using the same AtomSelector in multiple
ForceCalculations. Looking at equation \eqref{eq:forces} it can be seen that
in that scenario the same weight is used with both a positive and a negative
sign.

Testing is an important part of developing a simulation tool. Unit tests have
been written for all the main functionality of the Mixer. Specifically the
CalcBox weight generation and molecule entry/exit have been tested carefully.
Many of the classes support a debug level that can be specified in the
constructors. Debug output is written into a human readable textfile.
\subsection{Using The Mixer}
In this section we examine how the Mixer is used in practice. We prepare a
simple multi-scale simulation with just two hydrogen atoms.  The code snippets
demonstrate the procedure in a more tangible form.

There are four parts necessary for a Mixer calculation. First a valid ASE Atoms
object must be created, the second component is a set of ASE calculator objects,
third are the force and energy Mixer calculations and finally the fourth and
final piece are the AtomSelector objects that are used to carve the Atoms object
to parts given to each calculation.

The Atoms object can be created in the normal ASE way. Typically it is created
by a separate script that stores it in an ASE Trajectory file and read into
the actual calculation script later. Mixer places no restrictions to the
contents of the Atoms object. The only special preparation that needs to be
performed is executing Mixer.set\_atom\_ids(atoms) method, which adds a new
Atoms-internal array to the atoms object, containing unique integer
identifiers to each atom.

\lstset{basicstyle=\footnotesize, frame=single, captionpos=b, language=Python}
\lstset{caption=Declare Atoms object, label=code:atoms}
\begin{lstlisting}
atoms = Atoms("H2", position = [(0, 0, 0), (0, 0, 0.76470)],
              cell = (60, 60, 60))

Mixer.set_atom_ids(atoms)
\end{lstlisting}

This part is in accordance to the way Atoms class is
designed to work, so it does not interfere with any existing ASE functionality
and is automatically handled correctly by the various splicing and extending
methods of the Atoms class.  These unique identifiers are critical for
correctly combining forces provided by the sub-calculators.

After preparing the Atoms object with the unique IDs, the user can proceed to
creating instances of the ASE calculators to be used by the Mixer.

\lstset{caption=Create ASE calculators, label=code:calcs}
\begin{lstlisting}
gpaw_calc = GPAW(nbands=2, txt="gpaw.log")
reaxff_calc = ReaxFF(ff_file_path=get_datafile("ffield.reax"),
                     implementation="C")
\end{lstlisting}

In this simulation we want one of the hydrogen atoms to be in the classical
region, and the other in the quantum region. This is obviously quite
unphysical and the results are not very useful, but does serve well as a
straightforward example. We proceed by creating two selectors to split the atoms
into these two regions.

\lstset{caption=AtomSelector objects, label=code:atomselectors}
\begin{lstlisting}
full_system = AtomListSelector([0, 1],
                               {0: 1.0, 1: 1.0})
qm_region = AtomListSelector([1],
                             {1: 1.0})
\end{lstlisting}

The atom ids that the selector will select for are 0, 1 for the full system,
and 1 for the QM region. The second parameter to the constructor is the
set of weights assigned to each atom.

Now we are ready to define the Calculation objects for this system. Since we
are going to implement, at least in principle, the calculations based on
equations \eqref{eq:forces} and \eqref{eq:energies}, we need six calculation
objects. Three for energies and three for forces. This may look a little
verbose, but considering the flexibility this allows, the trade off is worth
it.
\lstset{caption=Calculation objects, label=code:calculations}
\begin{lstlisting}
forces_full_system_reaxff = ForceCalculation("forces_full",
                                             full_system)
forces_full_system_reaxff.calculator = reaxff_calc
forces_full_system_reaxff.cell = (60.0, 60.0, 60.0)

forces_qm_region_gpaw = ForceCalculation("forces_qm_gpaw",
                                         qm_region)
forces_qm_region_gpaw.calculator = gpaw_calc
forces_qm_region_gpaw.cell = (6.0, 6.0, 6.0)

forces_qm_region_reaxff = ForceCalculation("forces_qm_reaxff",
                                           qm_region)
forces_qm_region_reaxff.calculator = reaxff_calc
forces_qm_region_reaxff.cell = (60.0, 60.0, 60.0)
forces_qm_region_reaxff.coeff = -1.0


energy_full_system_reaxff = EnergyCalculation("energy_full",
                                              full_system)
energy_full_system_reaxff.calculator = reaxff_calc
energy_full_system_reaxff.cell = (60.0, 60.0, 60.0)

energy_qm_region_reaxff = EnergyCalculation("energy_qm_reaxff",
                                            qm_region)
energy_qm_region_reaxff.calculator = reaxff_calc
energy_qm_region_reaxff.cell = (60.0, 60.0, 60.0)
energy_qm_region_reaxff.coeff = -1.0

energy_qm_region_gpaw = EnergyCalculation("energy_qm_gpaw",
                                          qm_region)
energy_qm_region_gpaw.calculator = gpaw_calc
energy_qm_region_gpaw.cell = (6.0, 6.0, 6.0)
\end{lstlisting}

Now all that is left is creating the Mixer calculator object and running the
simulation. The Mixer constructor takes two lists that comprise the sets of
force and energy calculations.
\lstset{caption=Mixer instance creation, label=code:mixer}
\begin{lstlisting}
mixer = Mixer(forces=[forces_full_system_reaxff,
                      forces_qm_region_gpaw,
                      forces_qm_region_reaxff],
              energies=[energy_full_system_reaxff,
                        energy_qm_region_reaxff,
                        energy_qm_region_gpaw])
atoms.set_calculator(mixer)
\end{lstlisting}
From this point onwards the atoms object can be used in exactly the same way
as usual in ASE. All calculations will happen transparently in the background.
\subsection{Mixer Unit Tests}
\label{sec:mixerbox}
A rudimentary unit test set has been developed. The tests
have been built using the Python unittest class. Main focus is on the
AtomSelectors and the supporting OctreeNode. The tests exercise the classes
in isolation to make the results easier to interpret.

For OctreeNode the tests create and populate a volume and test finding
these objects using various search radius. This mimics the way CalcRegion
atom selector uses OctreeNode. AtomSelector tests focus on the CalcBox class
as CalcRegion and the underlying AtomSelector functionality gets completely
and predictably covered that way too. Transition region weight generation is
a key functionality and is extensively tested. Furthermore, the computations
demonstrated in section \ref{sec:computations} have been encapsulated into
test cases as well.

For the CalcBox class we have developed a separate mixer\_box.py Python script
that can be used to run a two calculator computation as described by equations
\eqref{eq:energies}, \eqref{eq:forces} and figure \ref{fig:calc_box_1}. This
script has been used during development for ad-hoc testing, but it is of
sufficient quality to be used by users. It takes a number of command line
parameters that control its behavior including specification of
ASE trajectory file for input, the dimensions of the calculation regions and
choosing the molecular dynamics to use and the number of steps to run among
other variables. It also serves as a complete example of how the Mixer framework
can be utilized to construct computation schemes for multiscale calculations.
%
%
\section{Simple molecular systems in hybrid simulations}\label{sec:computations}
\subsection{Interaction energy}\label{sec:demo}
\begin{figure}[!ht]
\centering
\includegraphics[width=0.3\textwidth]{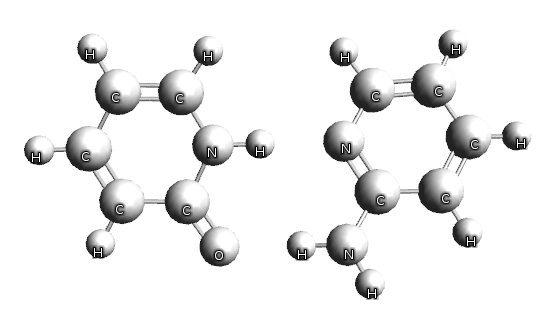}
\caption{2-pyridoxine 2-aminopyridine complex}
\label{fig:dimer_system}
\end{figure}

To demonstrate the use of the LAMMPS interface and the Mixer using the implemented
framework \cite{MultiASE} we calculate the 
interaction energy of a
2-pyridoxine 2-aminopyridine complex shown in figure \ref{fig:dimer_system}.
The equation for interaction energy of the dimer is
\begin{equation}\label{eq:dimer_interaction_energy}
H_{interaction} = H_{total} - H_{2-pyridoxine} - H_{2-aminopyridine}.
\end{equation}

In the Mixer calculations we treat the molecule on the left (2-pyridoxine) using
GPAW and the full system using LAMMPS/ReaxFF. In table \ref{tab:dimer_energies}
are presented the molecular and interaction energies computed first using ReaxFF
and GPAW alone, and then together through Mixer. The Mixer calculations have
been performed to match the energy equation \eqref{eq:energies} and in code look
very similar to what is found in listings \ref{code:atoms} through
\ref{code:mixer}.

\begin{table}[!ht]
\centering
\caption{2-pyridoxine 2-aminopyridine energies}
\label{tab:dimer_energies}
\begin{tabular}{lrrr}
\toprule
Energy & ReaxFF & GPAW/PBE & Mixer (GPAW + ReaxFF)\\
\midrule
2-pyridoxine & -62.339058 & -78.485885 & -78.519775\\
2-aminopyridine & -66.263778 & -83.629371 & -66.263778\\
Combined & -128.974632 & -162.745084 & -145.155333\\
Interaction & -0.371795 & -0.629827 & -0.371780\\
\bottomrule

\end{tabular}
\end{table}

The interaction energies found in table \ref{tab:dimer_energies} differ somewhat
from the value -0.72eV published in \cite{dimer_energy}, but this is
expected as that value has been achieved using coupled-cluster method, and
these here are of lesser accuracy by design.
\subsection{Forces}
In order to verify the force mixing functionality of the Mixer, a system
consisting of a methane molecule is used.
The molecule is moved along the x-axis so that it passes through the central
region of the simulation box. This motion on the x-axis is the only time
integration done on the system, the relative positions of atoms remain constant
to make comparison of forces calculated in different molecule positions easier.
The computations are configured similar to what is shown in figure
\ref{fig:calc_box_1} but expanded to three dimensions.

\begin{figure}[!ht]
\centering
\includegraphics[width=0.5\textwidth]{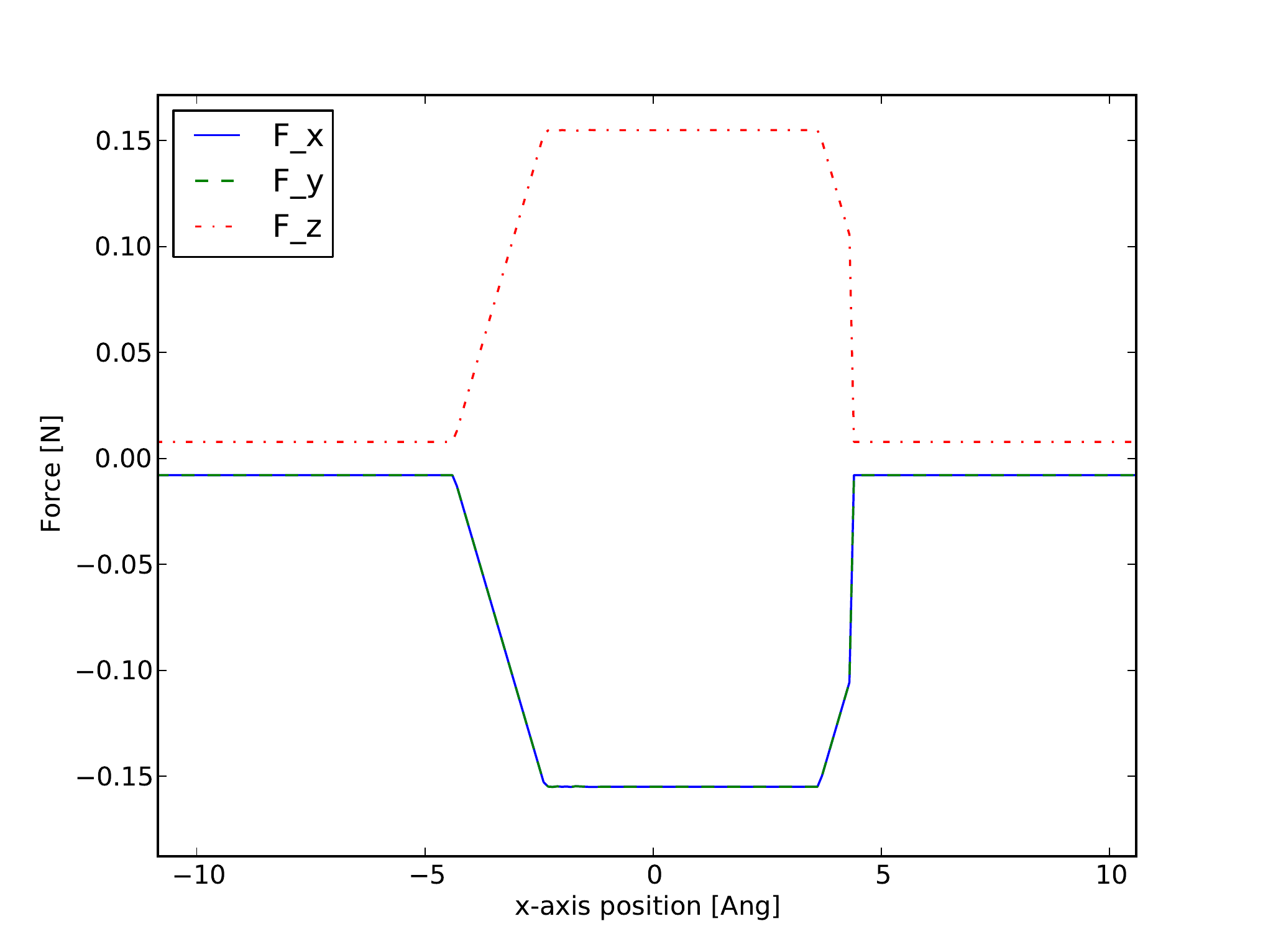}
\caption{Force components acting on a methane hydrogen atom}
\label{fig:force_components}
\end{figure}

%
LAMMPS/ReaxFF is used to model the full system using 100x100x100 angstrom cell
and a small central region of 10x10x10 angstroms is handled by GPAW. The central
quantum box has a 2 angstrom transition region where the mixing weights  change
from 0.0 to 1.0 while moving inwards. In addition the GPAW calculation is
performed with a 3 angstrom thick vacuum layer added around the actual quantum
box zone. The purpose of this experiment is to show how the intra-molecule
forces acting on the atoms change when the molecule enters and exits the quantum
box.
section \ref{sec:mixerbox}.
The calculations implement the force equation \eqref{eq:forces}. The force
components for a single hydrogen atom are shown in figure
\ref{fig:force_components} which also shows how the forces gradually change from
the values produced by ReaxFF to the ones computed by GPAW as the CH4 molecule
enters the quantum box from the left. While it stays within the GPAW box the
forces are constant, and then migrate back to ReaxFF values as it exits.
Worth noticing is how the entry is completely smooth, but the exit has a sharp
jump. This is caused by Mixer making sure that all atoms of a single molecule
are calculated with the same calculator. Now when the molecule is exiting, one
of its atoms moves outside the GPAW calculation cell while the one being
observed is still a significant distance from that boundary. At that point the
entire molecule must be excluded from the GPAW computation. In other words
the atom being inspected is on the trailing edge of the methane molecule.
This discontinuity can be made less severe by using a larger transition region.
\section{Conclusion}
We have created a force and energy mixer for ASE. 
The Mixer framework was used to perform an ONIOM-like multiscale
computation using GPAW and LAMMPS/ReaxFF and simple test scenarios were used to
verify both energy and force mixing functionality.
We have also created an ASE/LAMMPS interface designed to allow easy additions of new force fields. Current supported force fields include ReaxFF, CHARMM general force field, COMPASS and ClayFF.
The MultiASE framework has proven to be flexible and robust enough in these
simulations, and it is proposed here as a base for developing new and more advanced multiscale strategies. 
\section{Acknowledgments}
L.L and O.L-A acknowledge support by the Academy of Finland through its Centre of Excellence Program (project no. 251748). Computational resources were provided by Finland IT center for Science (CSC).
\bibliographystyle{is-abbrv}
\bibliography{bibliography}
\end{document}